\documentclass[a4paper,10pt]{article}

\usepackage{hyperref}

\title{A New Method for calculating Nonrelativistic Feynman Propagators in Momentum Space}
\author{Oem Trivedi \\ email id : oem.maapaa@gmail.com }
\date{\today}

\begin{document}
	 
	\maketitle
	
	\begin{abstract}
		
		In this paper, we present a new method for calculating Feynman Propagators in Momentum Space. The method is fundamentally based on Schwinger's Method for non relativistic propagators. While the aforementioned method has been used heavily in some papers to deduce various nonrelativistic propagators in position space, we show that on the basis of this method, we could developa technique to calculate nonrelativistic propagators in momentum space as well. We also show how to extract the Propagator for the concerned case for the harmonic oscillator.
	
\end{abstract}
keywords : Momentum space ; non relativistic propagators ; Schwinger's Method
\section{Introduction}
In recent papers \cite{aragao2007non}, \cite{barone2003three} Schwinger's oscillator method/Schwinger's Method for non relativistic propagators has been heavily applied to find out non relativistic propagators . Schwinger's method in it's essence, is using the Heisenberg representation and calculating Position and Momentum operators in it. Then ordering we order these operators in a certain way and substitute them in the Hamiltonian for the particle. The last step involves solving the supposedly easier path integral. In \cite[page 94]{kleinert2009path} and \cite[Page 100]{feynman2010quantum},it has been shown that path integral formulation of quantum mechanics can be done substantially well in the momentum space representation as well. Hence, in our paper, we attempt to formulate Schwinger's method for non relativistic propagators in the momentum space and calculate some elementary propagators using that method. We don't take into account the effects of quantum spin as we are interested in non relativistic systems. 

\section{Preliminaries}

Schwinger's method for calculating propagators is most prominently well known in Quantum Field Theory and was developed in it's original conception as a relativistic method. But as mentioned earlier, this has been used in some papers to be a nice way to calculate non relativistic propagators as well. We briefly review this method now. 

We know that operators in Heisenberg representation are defined to be time dependent as , \begin{equation}
\ \hat{O_H} (t) = e^\frac{iHt}{\hbar} \hat{O_s} e^\frac{-iHt}{\hbar} 
\end{equation}   , where $ \hat{O_H} $ and $ O_s $ are the operators in Heisenberg and Schrodinger Representations respectively.  And we also know that the differential equation for a Feynman Propagator going from (x,t) to $\ (x\prime,t\prime) $ is 
\begin{equation}
\i\hbar \frac{\partial}{\partial t} <x\prime,t\prime|x,t> = <x\prime,t\prime |\hat{H}|x,t>
\end{equation}
And also that \begin{equation}  \hat{X}(t\prime) |x\prime,t\prime> = x\prime |x\prime,t\prime> $$ $$\ \hat{X}(t) |x,t>= x |x,t> \end{equation}  Similar relations hold true for the momentum operators at corresponding times as well. 

With that cleared up we summarize the Schwinger Method for calculating nonrelativistic propagators as follows :-

\begin{enumerate}
	\item  We first compute the time evolution of the momentum and position operators through Heisenberg's equations of motion as \begin{equation}
	\frac{d\hat{P}}{dt} = \frac{i}{\hbar} [\hat{H},\hat{P}] $$ $$\ 
	\frac{d\hat{X}}{dt} =\frac{i}{\hbar} [\hat{H},\hat{X}] \end{equation}
	
	\item Then we use the solutions for $\hat{X} $ and $\hat{P} $ operators obtained in step 1 and write the Hamiltonian of the concerned problem in terms of these operators. We get rid of the P operators in the Hamiltonian and write everything in terms of $\hat{X}(t) $ and $\hat{X}(t\prime) $. We then order our Hamiltonian in such a way that we always have the X operators strictly in the order $ \hat{X}(t)\hat{X}(t\prime) $ , whenever they appear in multiplicative combinations. We could do this by using the commutator $ [\hat{X}(t), \hat{X}(t\prime)] $.  
	After doing that we evaluate the propagator as \begin{equation}
	 <x\prime,t\prime|x,t> = C(x\prime,x) e^{\frac{-i}{\hbar} \int_{t}^{t\prime} H(x\prime,x,t) dt} 
	\end{equation} 
	\item The only thing which is left to do is to figure out the Constant $ C(x\prime,x) $ This can be done by imposing certain conditions on the constant with use of the Eigenvalue equations for the momentum operator at $ t $ and $ t\prime $ . These are  \begin{equation}
	<x\prime,t\prime|\hat{P}(t)|x,t>=-i\hbar \frac{\partial}{\partial x\prime} <x\prime,t\prime|x,t>
	\end{equation}  And \begin{equation}
	 <x\prime,t\prime|\hat{P}(0)|x,t>=i\hbar \frac{\partial}{\partial x} <x\prime,t\prime|x,t> 
	\end{equation}
	   
\end{enumerate}
 
Now, if we analyze the time evolution of the momentum space state vector, we find that it evolves identically to the position space state vector. The time evolution operator works in the same way for momentum space state vectors as it does for the position space state vectors. This is clear from that fact that the Time dependent Schrodinger Equation in p-space \cite[Page 112, 4.117]{robinett1997quantum} is , \begin{equation}
 E\phi(p,t) = i\hbar \frac{\partial \phi (p,t)}{\partial t} \end{equation}  Which gives us the time evolution of  $\phi (p)$ as , \begin{equation}
  \phi(t)=A e^{\frac{-iHt}{\hbar}} 
 \end{equation} Where A is some constant in the simple case.

But when we take into account the path integral formulation, we see that analogous to the position space, we could define the propagator for momentum space state vectors as , \begin{equation}
\phi (p\prime,t\prime) = <p\prime,t\prime|p,t> \phi(p,t)
\end{equation} Where $ <p\prime,t\prime|p,t> $ is the propagator for the momentum space state vector from (p,t) to $ (p\prime,t\prime) $ . 

With all the preliminaries cleared up, we now devise the equivalent of Schwinger's method for non relativistic in position space propagators for non relativistic propagators in momentum space. 

\section{Schwinger's method for Non Relativistic Feynman Propagators in momentum space }

This method quite similar to that for the  position space, with just a few changes. To find the Feynman propagator for a non relativistic particles in momentum space, we follow the following steps :-

\begin{enumerate}
	\item We write the Heisenberg equations for the $\hat{X}$ and $\hat{P}$ just as we did in the initial step of the method for position space.
	\item Then we proceed to construct the Hamiltonian for the concerned problem in terms of $\hat{X} $ and $\hat{P}$ operators. After doing this, we proceed to order the Hamiltonian. Now recall that in the method for position space we ordered the Hamiltonian by writing out the $\hat{P} $ operators in terms of the $\hat{X}$ and then proceeding to write out any multiplicative combination of the $\hat{X}$ operators from right to left going forward in time (Which we wrote as $\hat{X}(t\prime) \hat{X}(t) $ because we were dealing with time evolution in the case of $ t=t $ to $ t=t\prime $ ). Here we would order the Hamiltonian by writing the $\hat{X}$ operators in terms of the $\hat{P}$ operators. And then we shall, like in position space, write any multiplicative combination of $\hat{P}$  operators from right to left going, as we want to proceed forward in time as $\hat{P}(t\prime) \hat{P}(t) $ by using $ [\hat{P}(t),\hat{P}(t\prime)] $ where $ t\prime > t $ . 
	\item Now the Propogator comes in the form \begin{equation} <p\prime,t\prime|p,t> = C(p\prime,p) e^{\frac{-i}{\hbar} \int_{t}^{t\prime} H(p\prime,p,t) dt } \end{equation}
	The constant in the propagator can be determined by imposing conditions on them  which we obtain from the position operator in momentum space as, \begin{equation}
	 <p\prime,t\prime |\hat{X}(t\prime)|p,t>=i\hbar \frac{\partial}{\partial p\prime} <p\prime,t\prime|p,t> 
	\end{equation} And \begin{equation}
	 <p\prime,t\prime|\hat{X}(t)|p,t>=-i\hbar \frac{\partial}{\partial p} <p\prime,t\prime|p,t> 
	\end{equation} 
\end{enumerate}

And hence, in this manner we could calculate the non relativistic propagators for momentum space state vectors using Schwinger's method. It's evident that it is very much similar to the actual method for the position space state vectors just with some changes. Now we illustrate this method by taking the example of the Simple Harmonic Oscillator. 

\section{Example - The Harmonic Oscillator}

In this section, we illustrate the method discussed in section 3 for the particular case of the harmonic oscillator. We consider the system to be evolving from a time $ t=0 $ to a time $ t=t $ for simplicity. Now, we first calculate the time evolution of the $\hat{X}$ and $\hat{P}$ operators. This is pretty straightforward as the Harmonic Oscillator is a standard exercise in quantum physics textbooks. So we have for the $\hat{X}$, \begin{equation}
 \frac{d\hat{X}}{dt}=\frac{i}{\hbar}[\hat{H},\hat{P}]=\frac{\hat{P}}{m} 
\end{equation} and for the $\hat{P}$ operator we have, \begin{equation}
\frac{d\hat{P}}{dt}=\frac{i}{\hbar}[\hat{H},\hat{P}]=-m\omega^2\hat{X} \end{equation} The well known solutions of these equations are, \begin{equation}
  \hat{X}(t)= \hat{X}(0)\cos(\omega t) + \frac{\hat{P}(0)}{m\omega} \sin(\omega t) 
\end{equation} And the Momentum operator is , \begin{equation}
 \hat{P}(t)=-m\omega \hat{X}(0) \sin(\omega t)+\hat{P}(0) \cos(\omega t) 
\end{equation}

The Hamiltonian for the harmonic oscillator is defined in the usual way. As the system is autonomous in time, the Hamiltonian is obviously not explicitly dependent on time. So that means $ \hat{H} (t) = \hat{H}(0) $. To illustrate this more clearly, we write that \begin{equation}
\hat{H} (t) = \frac{\hat{P}(t)^2}{2m} + \frac{m\omega \hat{X}(t)^2}{2} = \frac{\hat{P}(0)^2}{2m} + \frac{m\omega \hat{X}(0)^2}{2} = \hat{H}(0) 
\end{equation}

As it would be important for ordering the Hamiltonian, we note that at any time  $ t $ , the commutator \begin{equation}
 [\hat{P}(0),\hat{P}(t)]=m\omega i \hbar \sin(\omega t) 
\end{equation}. (19) allows us to write \begin{equation}
 \hat{P}(0)\hat{P}(t) = \hat{P}(t)\hat{P}(0) - m\omega i \hbar \sin (\omega t) 
\end{equation} With our Position operators ordered correctly, now we need to use them in the Hamiltonian. We choose to work with the Hamiltonian at time t for our purposes here. Then, after substituting the solutions for the harmonic oscillator equations in the Heisenberg picture and ordering the Hamiltonian using the commutation relation of the momentum operators, we get the Hamiltonian, \begin{equation}
 H(t)= \frac{\frac{P(t)^2 + P(0)^2 cos^2 (\omega t) + P(0)^2 \cot^2 (\omega t) + P(t)^2 \csc^2 (\omega t)}{2} + P(t) P(0) \csc (\omega t) \cot (\omega t) + i\omega \hbar \csc (\omega t) \cot (\omega t)}{m} 
\end{equation}

Putting this all into the propagator we defined earlier, we get the propagator in momentum space as \begin{equation}
 <p(t), t | p(0),0 > = C(p(t),p(0) )e^{\int_{0}^{t}\frac{-i H(t) dt}{\hbar}} 
\end{equation}

Substituting the hamiltonian we have the Propagator as, \begin{equation}
 <p(t), t | p(0),0 > = C(p(t),p(0) \exp^{\int_{0}^{t}\frac{-i}{m\hbar}( {\frac{P(t)^2 + P(0)^2 cos^2 (\omega t) + P(0)^2 \cot^2 (\omega t) + P(t)^2 \csc^2 (\omega t)}{2} + P(t) P(0) \csc (\omega t) \cot (\omega t) + i\omega \hbar \csc (\omega t) \cot (\omega t)})} 
\end{equation} Here we have just substituted the Hamiltonian completely into the integral.

This integral can be evaluated to find out that, \begin{equation}
 <p(t), t | p(0),0 > =  \frac{C (P(t), P(0)}{\sqrt{\sin (\omega t)}} \exp^{\frac{i( (P(t)^2 + P(0)^2 ) \cos (\omega t)- 2 P(t)P(0) }{2m\omega \sin (\omega t) \hbar}} 
\end{equation}

Now we can impose a condition on the propagator given how that we are dealing time  evolution from t=0 to t=t . It is that, \begin{equation}
 \lim\limits_{t\to 0 } <P(t),t|P(0),0> = \delta (P(t)-P(0) 
\end{equation} where $ \delta (x) $ is the usual Dirac delta function. 

We again, have, \begin{equation}
 <P(t),t|X(t)|P(0),0> = i\hbar \frac{\partial <P(t),t|P(0),0>}{\partial P(t)} $$ $$\ <P(t),t|X(0)|P(0),0>=-i\hbar \frac{\partial <P(t),t|P(0),0}{\partial P(0)} 
\end{equation}  

Using these conditions we find that, similar to \cite[equation 32, page 6]{barone2003three}, \begin{equation}
 \frac{\partial C(P(t),P(0)}{\partial P(t)} = 0 $$  and $$ \frac{\partial C(P(t),P(0)}{\partial P(0)} = 0 
\end{equation} And analogous to \cite[equation 33, page 6]{barone2003three}, we take \begin{equation}
 \lim\limits_{t\to 0 }<P(t),t|P(0),0> = \lim\limits_{t\to 0} \frac{C}{\sqrt{\sin (\omega t)}} \exp^{\frac{i [(P(t)^2 + P(0)^2) \cos (\omega t) - 2 P(t) P(0) ]}{2 m \omega \hbar \sin (\omega t)}} 
\end{equation}. Now if we apply the conditions that we have discussed for the propagator and normalize it, we find that $$ C= \sqrt{\frac{h}{i\omega \sin (\omega t)}} $$, where $ h $ is obviously the usual Planck's constant (and not the reduced Planck's constant $ \hbar = \frac{h}{2\pi} $ )  So we have our propagator as,\begin{equation}
  <P(t),t|P(0),0> = \sqrt{\frac{h}{i\omega \sin (\omega t)}} \exp^{\frac{i( (P(t)^2 + P(0)^2 ) \cos (\omega t)- 2 P(t)P(0) }{2m\omega \sin (\omega t) \hbar}} 
\end{equation} This is in agreement with the propagator obtained in the D=1 case in \cite[2.189]{kleinert2009path}. Hence we have demonstrated how our method can be used to calculate momentum space propagators. 

\section{Conclusion}

In conclusion, we have shown that Schwinger's method used for calculating Non relativistic Feynman propagators in position space can be used to calculate non relativistic Feynman propagators in momentum space. This can be done by slightly altering the method and making it more focused upon similar ideas in momentum space as in position space. To illustrate the method, we used the example of Harmonic Oscillator and deduced the propagator. 

\section{Acknowledgments}

I would like to thank Professor Andre Grossardt ( Department of Physics, Queen's University, Belfast, Northern Ireland) for endorsing me for quantum physics at arxiv. I would also like to thank Professor Chigak Itoi (Department of Physics , Nihon University, Tokyo, Japan ) and Professor Mario Rocca (Department of Physics, National University of La Plata, La Plata, Buenos Aires, Argentina) for guidance and helpful insights. I would also specially like to thank Professor Itoi for pointing towards the dirac delta initial condition on the propagator. 

\bibliographystyle{unsrt}
\bibliography{ShreeHari}

\begin{thebibliography}{1}

\bibitem{aragao2007non}
A~Aragao, H~Boschi-Filho, C~Farina, and FA~Barone.
\newblock Non-relativistic propagators via schwinger's method.
\newblock {\em Brazilian Journal of Physics}, 37(4):1260--1268, 2007.

\bibitem{barone2003three}
FA~Barone, Henrique Boschi-Filho, and C~Farina.
\newblock Three methods for calculating the feynman propagator.
\newblock {\em American Journal of Physics}, 71(5):483--491, 2003.

\bibitem{kleinert2009path}
Hagen Kleinert.
\newblock {\em Path integrals in quantum mechanics, statistics, polymer
  physics, and financial markets}.
\newblock World scientific, 2009.

\bibitem{feynman2010quantum}
Richard~P Feynman, Albert~R Hibbs, and Daniel~F Styer.
\newblock {\em Quantum mechanics and path integrals}.
\newblock Courier Corporation, 2010.

\bibitem{robinett1997quantum}
Richard~Wallace Robinett and RD~Murphy.
\newblock {\em Quantum mechanics: classical results, modern systems, and
  visualized examples}.
\newblock Oxford University Press New York, 1997.

\end{thebibliography}

\end{document}